\documentclass[twocolumn]{aastex62}

\usepackage{amssymb, amsfonts, amsmath,framed}

\usepackage{bm}
\expandafter\ifx\csname package@font\endcsname\relax\else
 \expandafter\expandafter
 \expandafter\usepackage
 \expandafter\expandafter
 \expandafter{\csname package@font\endcsname}
\fi
\expandafter\ifx\csname package@font\endcsname\relax\else
 \expandafter\expandafter
 \expandafter\usepackage
 \expandafter\expandafter
 \expandafter{\csname package@font\endcsname}
\fi
\def\bq{\begin{equation}}
\def\eq{\end{equation}}
\def\bqy{\begin{eqnarray}}
\def\eqy{\end{eqnarray}}

\begin{document}
\title{\large{\textbf{Magnetohydrodynamic Turbulence in the Plasmoid-Mediated Regime}}}

\correspondingauthor{Luca Comisso}
\email{lcomisso@princeton.edu}

\author{L. Comisso}
\affiliation{Department of Astrophysical Sciences, Princeton University, Princeton, New Jersey 08544, USA}
\affiliation{Princeton Plasma Physics Laboratory, Princeton University, Princeton, New Jersey 08543, USA}
\affiliation{Princeton Center for Heliophysics, Princeton University, Princeton, New Jersey 08543, USA}

\author{Y.-M. Huang}
\affiliation{Department of Astrophysical Sciences, Princeton University, Princeton, New Jersey 08544, USA}
\affiliation{Princeton Plasma Physics Laboratory, Princeton University, Princeton, New Jersey 08543, USA}
\affiliation{Princeton Center for Heliophysics, Princeton University, Princeton, New Jersey 08543, USA}

\author{M. Lingam}
\affiliation{Institute for Theory and Computation, Harvard University, Cambridge, Massachusetts 02138, USA}
\affiliation{Harvard-Smithsonian Center for Astrophysics, Cambridge, Massachusetts 02138, USA}

\author{E. Hirvijoki}
\affiliation{Princeton Plasma Physics Laboratory, Princeton University, Princeton, New Jersey 08543, USA}

\author{A. Bhattacharjee}
\affiliation{Department of Astrophysical Sciences, Princeton University, Princeton, New Jersey 08544, USA}
\affiliation{Princeton Plasma Physics Laboratory, Princeton University, Princeton, New Jersey 08543, USA}
\affiliation{Princeton Center for Heliophysics, Princeton University, Princeton, New Jersey 08543, USA}

\begin{abstract}
Magnetohydrodynamic turbulence and magnetic reconnection are ubiquitous in astrophysical environments. In most situations, these processes do not occur in isolation, but interact with each other. This renders a comprehensive theory of these processes highly challenging. Here, we propose a theory of magnetohydrodynamic turbulence driven at large scale that self-consistently accounts for the mutual interplay with magnetic reconnection occurring at smaller scales. Magnetic reconnection produces plasmoids that grow from turbulence-generated noise and eventually disrupt the sheet-like structures in which they are born. The disruption of these structures leads to a modification of the turbulent energy cascade, which, in turn, exerts a feedback effect on the plasmoid formation via the turbulence-generated noise. The energy spectrum in this plasmoid-mediated range steepens relative to the standard inertial range and does not follow a simple power law. As a result of the complex interplay between turbulence and reconnection, we also find that the length scale which marks the beginning of the plasmoid-mediated range and the dissipation length scale do not obey true power laws. The transitional magnetic Reynolds number above which the plasmoid formation becomes statistically significant enough to affect the turbulent cascade is fairly modest, implying that plasmoids are expected to modify the turbulent path to dissipation in many astrophysical systems.

$\,$

$\,$

\end{abstract}

\section{Introduction} \label{SecIntro}

Magnetohydrodynamic (MHD) turbulence plays an essential role in a variety of space and astrophysical systems, ranging from stellar coronae \citep{Matthaeus99,Cranmer2007} and black hole accretion disks \citep{BalbusHawley98,BrandSubra05}, to the interstellar medium \citep{Armstrong95,LithGold01} and galaxy clusters \citep{Zweibel97,Subra06}. It is therefore of paramount importance to understand MHD turbulence at a fundamental level to arrive at a detailed comprehension of these phenomena. 

A characteristic feature of MHD turbulence is~the development of small-scale current sheet structures that are prone to magnetic reconnection \citep[e.g.,][]{ML86,Politano89,Biskamp2003,Servidio09,Zhdankin13,Wan2013}. Indeed, it is well known that thin current sheets can be unstable to reconnection instabilities \citep{FKR_1963,Coppi1976}. This, of course, raises the important issue of evaluating whether, and how, magnetic reconnection impacts the turbulent cascade. Magnetic reconnection via tearing modes does indeed produce plasmoids (flux ropes) within current sheets that give rise to a turbulent scenario which is qualitatively different from the homogeneous Alfv\'enic turbulence picture \citep{HuBha16}. 

For plasmoids to be relevant, they have to disrupt the current sheet within its characteristic lifetime, i.e. within one nonlinear eddy turnover time $\tau_{\rm{nl}}$. On this basis, it has been proposed \citep{Carbone1990,Mallet2017,LouBold2017,BoldLou2017} that current sheet structures in a turbulent environment disrupt when $\gamma \tau_{\rm{nl}} \sim 1$, with $\gamma$ indicating the growth rate of the fastest tearing mode. This condition has been used to evaluate the scale at which the inertial range breaks and to propose a different energy spectrum and dissipation scale, thereby providing an intuitive picture of plasmoid effects in a strong turbulent cascade. On the other hand, assuming $\gamma \tau_{\rm{nl}} \sim 1$ at disruption overestimates the effects of magnetic reconnection on the turbulent cascade. In fact, for this case, the seed noise of the plasmoid instability would be amplified only by a factor $\sim e^1 \approx 2.7$ in one eddy turnover time, which is too small to destroy a current sheet structure. Consequently, current sheet disruption occurs at smaller scales when $\gamma \tau_{\rm{nl}} \gg 1$ \citep{Comisso2016,Comisso2017,Huang2017} \footnote{Furthermore, a technical but important aspect is that $\gamma \gg 1/\tau_{\rm{nl}}$ is necessary for applying standard tearing mode theory.}. In particular, $\gamma \tau_{\rm{nl}}$ at disruption must depend on both the noise level and the magnetic diffusivity $\eta$.
It follows that the action of the plasmoid instability is dictated by the turbulence itself, and a more refined quantitative analysis necessitates a detailed theoretical treatment that incorporates this complex interplay.

It is the primary goal of this paper to develop a refined quantitative theory that self-consistently accounts for the mutual interaction of magnetic reconnection and turbulence. Fluctuations arising from turbulence provide the noise that seeds plasmoid growth, and, in turn, the plasmoids disrupt current sheet structures, thereby modifying the turbulent energy cascade and eventually exerting a feedback effect on the plasmoid instability via turbulence-generated noise. This unified picture is a novel feature of our analysis, and is manifested in the relevant physical quantities that are no longer simple power laws.

\section{Current sheet disruption} \label{SecCSdisr}

An important feature of MHD turbulence is that it becomes increasingly anisotropic toward small scales within the inertial range.
In particular, turbulent structures display anisotropy in all three directions \citep{Politano95,Zhdankin13,Makwana2015}: $\lambda \ll \xi \ll l_\parallel$. Here, $\lambda$ and $\xi$ are the dimensions of a given structure in the plane normal to the local guiding magnetic field, the field-perpendicular eddy size and the fluctuation-direction scale, respectively, while $l_\parallel$ is the dimension of the structure along the magnetic field. This anisotropy can be understood in terms of two phenomenological observations, the critical balance condition \citep{GS95} and the scale-dependent alignment \citep{Boldyrev2006,Chandran15,MalletScheko17}. The first expresses the tendency of the turbulence dynamics to be attracted toward a state in which $\tau_A={l_\parallel}/{V_A} \sim \tau_{\rm{nl}}$ throughout the inertial range of the strong cascade, while the latter is based on the realization that the magnetic and velocity fluctuations, $\delta {v_\lambda }$ and $\delta {b_\lambda}$, become spontaneously aligned in the field-perpendicular plane with a small scale-dependent angle $\theta_\lambda$. In particular, the inverse-aspect-ratio ${\lambda}/{\xi} \sim \sin {\theta_\lambda} \ll 1$ can be estimated as $\sin {\theta _\lambda } \simeq {\theta _\lambda } \sim {\delta {b_\lambda}}/{V_A} \sim {\left( {\lambda }/{L} \right)^{1/4}}$ \citep{Boldyrev2006}, where $V_A$ is the Alfv\'en velocity based on the background magnetic field $B_0$ and $L$ is the (perpendicular) outer-scale, at which turbulence is assumed to be critically balanced. Note that uniform plasma density is considered and Alfv\'enic units are used, i.e. $V_A=B_0$.
 
The fact that turbulent fluctuations in the inertial range are characterized by $\lambda \ll \xi \ll l_\parallel$ implies that magnetic field fluctuations give rise to effective current sheets of thickness $\lambda$ and length $\xi$. However, very large aspect ratios $\xi/\lambda$ cannot be sustained against the plasmoid instability. The aspect ratio at which the plasmoid instability becomes critical can be calculated from simple first principles. It can be done by considering two different viewpoints, that of a rapidly forming current sheet, and that of a static one. The solution for a generic current sheet formation, based on a principle of least time, has been derived in \citet{Comisso2016,Comisso2017}. For Alfv{\'e}nic (exponentially shrinking) current sheet formation, it was shown that plasmoids become nonlinear when
\begin{equation}\label{espl_res_a_GENERAL} 
\frac{\lambda_*}{\xi_*} \simeq S_\xi ^{-1/3}  {\left[ {\ln \left( {\frac{{S_\xi ^{ - 3(2 + \alpha )/4}}}{{{2^6}{\hat \epsilon ^3}}}{{\left( {\frac{{{\lambda _*}}}{{{\xi _*}}}} \right)}^{\! 3(2 - 5\alpha )/4}}} \right)} \right]^{-2/3}}
  \, ,
\end{equation} 
where $S_\xi = \xi {v_{A\lambda}}/\eta$ is the Lundquist number based on $\xi$ and ${v_{A\lambda}}$, which is the Alfv\'en speed associated with the perturbed magnetic field at scale $\lambda$. Moreover, $\hat \epsilon = \epsilon/(\delta {b_\lambda }\xi)$ is a normalized amplitude of the noise that seeds the instability, and $\alpha$ is an index that depends on the spectrum of the noise. \footnote{From Eq. (\ref{espl_res_a_GENERAL}) it should be clear that $S_\xi$ and $\hat \epsilon$ must be understood as being evaluated at the scale indicated by the asterisk.} 
Here the noise is assumed to have a general power-law form, i.e. ${\psi}_{0} = {\epsilon} {{({k_\xi}{\xi})}^{- \alpha}}$, with $k_\xi$ indicating the wavenumber in the $\xi$-direction, which turns out to be valid as a zeroth order approximation. 

On the other hand, an alternative estimation of ${\lambda_*}/{\xi_*}$ can be done by determining the growth rate of the instability in a fixed current sheet such that the amplitude of the perturbation grows from the noise level to nonlinearity ($\delta_{\rm{in}}$) in one $\tau_{\rm{nl}}$, i.e.
\begin{equation} \label{Condition_gamma}
\ln \left( {\frac{\delta_{\rm{in}}}{w_0}} \right) = \ln \left[ {\frac{{{{\left( {k_\xi \xi } \right)}^{(\alpha  - 1)/2}}}}{{2{{\hat \epsilon }^{1/2}}}}{\left( {\frac{{\gamma {\tau _{A\xi }}}}{{{S_\xi }}}} \right)^{\! 1/4}}} \right] = \frac{\gamma \tau_{\rm{nl}}}{2} \, .
\end{equation}
Here we have used $\delta_{\rm{in}}= {\big[ { \eta {\gamma}{\lambda^2}/{(k_\xi{v_{A\lambda}})^2} } \big]^{1/4}}$ for the inner resistive layer width and $w_0 = 2{\left( {{\psi _0}\lambda /\delta {b_\lambda}} \right)^{1/2}}$ for the seed geometrical width \citep[e.g.,][]{Biskamp2003}.
Since the current sheet has no time dependence in this case, it is clear that the mode that disrupts the sheet is the fastest growing mode, for which
\begin{equation} \label{gamma_max} 
\gamma_{f} \simeq c_{\gamma f} \; \frac{v_{A\lambda}}{S_\lambda^{1/2} \lambda} = S_\xi^{-1/2} \frac{c_{\gamma f}}{\tau_{A\xi}} {\left( {\frac{\xi}{\lambda}} \right)^{\! 3/2}} \, ,
\end{equation}
\begin{equation} \label{k_max} 
k_{\xi f}  \simeq c_{kf} \; \frac{1}{S_\lambda^{1/4} \lambda} = \frac{c_{kf}}{  {S_\xi ^{1/4}}\xi}{\left( {\frac{\xi}{\lambda}} \right)^{\! 5/4}}  \, ,
\end{equation}
where $S_\lambda = \lambda {v_{A\lambda}}/\eta$ and $\tau _{A\xi} = \xi /{v_{A\lambda}}$, while the multiplicative coefficients are $c_{\gamma f} \approx 0.623$ and $c_{kf} \approx 1.358$ \citep{FKR_1963,Coppi1976} for the common Harris-type sheet. Furthermore, taking the local 3D anisotropy into account, one can define the nonlinear timescale in the inertial range as $\tau_{\rm{nl}} = \lambda/({\delta {v_\lambda }\sin {\theta_\lambda}})$, and from critical balance $\tau_{\rm{nl}} \sim {(\lambda L)^{1/2}}/{V_A} \sim \tau_{A\xi}$. Therefore, replacing $\tau_{\rm{nl}}$ with $\tau _{A\xi}$ and substituting Eqs. (\ref{gamma_max}) and (\ref{k_max}) into Eq. (\ref{Condition_gamma}), we end up with an equation identical to Eq. (\ref{espl_res_a_GENERAL}) up to a multiplicative constant of 1.52. \footnote{This agreement arises because the intrinsic timescale of the plasmoid instability is near-universal for exponentially thinning current sheets \citep{Comisso2016,Comisso2017}.}  

Equation (\ref{espl_res_a_GENERAL}) can be solved exactly in terms of the Lambert $W$ function to obtain
\begin{equation}\label{espl_res_a} 
\frac{\lambda_*}{\xi_*} \simeq  {\left( {\frac{\underline \alpha }{{S_\xi ^{1/2} \, W(\zeta )}}} \right)^{\! 2/3}}   \, ,
\end{equation}
where we have introduced 
\begin{equation}\label{} 
\zeta = \underline \alpha \, {\left( {{2^6}{\hat \epsilon ^3}} \right)^{\!- \underline \alpha} } \, S_\xi ^{(\alpha  - 4)/(2 - 5\alpha )} \, ,
\end{equation}
with ${\underline \alpha} = 2/(2-5\alpha)$. 
Given the inverse-aspect-ratio ${\lambda_*}/{\xi_*}$, we can easily show that the growth rate at the end of the linear phase is 
\begin{equation}\label{esplicit_gamma} 
{\gamma _*}{\tau _{A\xi }} \simeq c_\gamma \frac{W(\zeta)}{{\underline \alpha}}  \, ,
\end{equation}
where $c_\gamma/c_{\gamma f} \approx 1$.
It can be shown \emph{a posteriori} that 
$\gamma_* \gg 1/\tau_{\rm{nl}}$, as it is required for the instability to amplify the perturbation to a significant size within one eddy turnover time. 

Note that the plasmoid width at the beginning of the nonlinear phase satisfies ${{\Delta}'} {{w }_*} \approx 2$ \citep{ComGra16,Comisso2017}, where ${\Delta}'$ is the tearing stability parameter associated with the wavelength that emerges first from the linear phase. This condition implies that the early nonlinear growth of the plasmoids occurs through a fast Waelbroeck phase \citep{Waelbr1989}, which could be further accelerated by the disruption of secondary current sheets, until the plasmoids reach the size $w \sim \lambda$ in a short timescale \citep{ComGra16}. However, despite this scenario, one can define the current sheet as having been already disrupted at the inverse-aspect-ratio ${\lambda_*}/{\xi_*}$. Indeed, the current density fluctuations caused by the plasmoids at the end of the linear phase are on the same order of the current density of the current sheet, implying that the latter has lost its integrity \citep{Huang2017}. 

The length scale at which the plasmoid instability can disrupt the current sheet structures is expressible in terms of the magnetic Reynolds number $R_m = {V_0}L/\eta$, which is defined with the outer-scale velocity $V_0$. Substituting the scale-dependent alignment relations ${\xi_*} \sim {\lambda_*}/{\sin {\theta_\lambda}}$ and  ${S_\xi }  \sim \left( {{\lambda _*}/L} \right){R_m}$ into Eq. (\ref{espl_res_a}), and then solving for $\lambda_*$, we obtain
\begin{equation}\label{lambda_*} 
\frac{{{\lambda _*}}}{L} \sim {\left( {\frac{{\bar \alpha }}{{R_m^{1/2} \,W(\chi )}}} \right)^{\! 8/7}} \, ,
\end{equation}
where we have defined
\begin{equation}\label{} 
\chi = \bar \alpha \, {\left( {{2^6}{\hat \epsilon^3}} \right)^{\! - \bar \alpha }}  R_m^{(4-\alpha)/(6 + 9\alpha)} \, ,
\end{equation}
with $\bar \alpha = -14/[9(2+3\alpha)]$. 
This relation yields the scale at which the plasmoid generation in tearing unstable current structures is statistically significant enough to affect the turbulent cascade. In particular, from Eq. (\ref{lambda_*}) we get the wavenumber $k_* = 2 \pi /{\lambda _*}$ above which the energy spectrum may change because the turbulent cascade enters the ``plasmoid-mediated range''. 

At this stage of the analysis, $\lambda_*$ still depends on the normalized amplitude $\hat \epsilon$ and the spectral index $\alpha$ of the noise seeding the plasmoid instability. Both of these quantities will be determined self-consistently in Sec. \ref{SecSpectrum} using information about the turbulence energy spectrum. 
Furthermore, a final important point that must be kept in mind is that the plasmoid instability can affect the turbulent cascade only if $\lambda_* \gg {\lambda_\eta}$, with ${\lambda_\eta}$ indicating the dissipation cutoff scale in the absence of statistically significant plasmoids. This condition, which sets a threshold for the magnetic Reynolds number $R_m$, is discussed in Sec. \ref{SecReyn}.

\section{Energy spectrum in the plasmoid-mediated range}\label{SecSpectrum}

In the inertial range of strong incompressible MHD turbulence, two field-perpendicular energy spectra, $E({k_ \bot }) \sim {\varepsilon^{2/3}} k_\bot^{-5/3}$ \citep{GS95} and $E({k_ \bot }) \sim {\varepsilon ^{2/3}}{L^{1/6}} k_\bot^{-3/2}$ \citep{Boldyrev2006}, have been derived under different assumptions. However, neither of these spectra can hold in the ``plasmoid-mediated range'', where the interactions due to the plasmoids become significant. To determine the energy spectrum in this range, we consider the standard constant energy flux requirement  
\begin{equation}\label{constant_energy_flux} 
\frac{{(\delta {b_\lambda })}^2}{{{\tau _{{\rm{nl}}}}}} = \mathrm{const} = \varepsilon  \, ,
\end{equation} 
but in this case it is the plasmoid instability that sets the nonlinear timescale and modifies the attainable aspect ratio of the current sheet structures. Therefore, using Eq. (\ref{esplicit_gamma}), we have
\begin{equation}\label{NL_time} 
{\tau _{{\rm{nl}}}} \simeq \frac{{{\lambda ^{3/2}}}}{{{{(\delta {b_\lambda }\eta )}^{1/2}}}}\frac{{W(\zeta )}}{\underline \alpha}  \, .
\end{equation}
Substituting this timescale into Eq. (\ref{constant_energy_flux}), we find that the magnetic field fluctuation at scale $\lambda$ satisfies the equation
\begin{equation}\label{pert_b_1} 
\delta {b_\lambda} \simeq \frac{{{\varepsilon^{2/5}}{\lambda^{3/5}}}}{\eta^{1/5}}{\left[ \frac{1}{\underline \alpha} { W \! \left( {{\underline \alpha} {{\left( {\frac{1}{{{2^6}{\hat \epsilon ^3}}}} \right)}^{\! \underline \alpha}}{{\left( {\frac{{\xi \delta {b_\lambda }}}{\eta }} \right)}^{\!{\frac{{\alpha  - 4}}{{2 - 5\alpha }}}}}} \right)} \right]^{2/5}} \, .
\end{equation}
To obtain ${\delta {b_\lambda }}$ as a function of $\lambda$ in the plasmoid-mediated range, we need to solve this implicit equation. This requires us to adopt a suitable expression for $\xi$ as a function of $\lambda$, as well as determine $\alpha$ and $\hat \epsilon$. These quantities can be obtained through an iterative procedure. However, since they occur in Eq. (\ref{pert_b_1}) through the Lambert $W$ function, which makes their dependence weak, we find that a single iteration is sufficient to accurately determine ${\delta {b_\lambda }}$. 

A simple approximation for $\xi$ can be readily obtained by neglecting the factors involving the Lambert $W$ function in the formulae that quantify the current sheet disruption. In this case $\sin {\theta _\lambda } \sim \lambda /(\delta {b_\lambda }{\tau _{{\rm{nl}}}}) \sim {\left( {{\lambda_*}/L} \right)^{1/4}}{\left( {\lambda/{\lambda_*}} \right)^{ -4/5}}$, which implies $\xi \sim L{\left( {{\lambda_*}/L} \right)^{3/4}}{\left( {\lambda/{\lambda_*}} \right)^{9/5}}$ in Eq. (\ref{pert_b_1}). On the other hand, the evaluation of $\alpha$ requires additional information from the energy spectrum. In particular, we are interested in fluctuations in the $\xi$ direction, since these are the perturbations that trigger tearing modes. Eq. (\ref{pert_b_1}) indicates that $\delta {b_\lambda} \propto  \lambda^{3/5}$ at the zeroth order. Therefore, since $\xi  \propto  \lambda^{9/5}$, we have $\delta {b_\lambda} \propto k_\xi ^{-1/3}$, which implies $E({k_\xi }) \propto k_\xi ^{-5/3}$ at the zeroth order. Finally, the relation $E({k_\xi }) \sim \psi_0^2{k_\xi } = ({\epsilon^2}/\xi){({k_\xi }\xi )^{ 1- 2\alpha }}$ allows us to specify
\begin{equation}\label{} 
\alpha = 4/3 \, , \quad {\underline \alpha} = -3/7 \, , \quad  \bar \alpha = -7/27 \, .
\end{equation}
Using this information, and the relation $L=V_A^3/\varepsilon$ from the constant energy flux requirement at the outer-scale, we can write the solution of Eq. (\ref{pert_b_1}) as
\begin{equation}\label{pert_b_2} 
\delta {b_\lambda } \simeq \frac{\varepsilon^{2/5}{\lambda ^{3/5}}}{\eta^{1/5}}{\left[  -\frac{9}{5} W_{-1} {\left( { - \frac{5}{9} \frac{{(2^2 \hat \epsilon)}^{5/3}}{R_m^{-8/9}} \frac{{{\lambda _*}}}{L}{{\left( {\frac{\lambda }{{{\lambda _*}}}} \right)}^{\!16/9}}} \right)} \right]^{2/5}} \, ,
\end{equation}
where $W_{-1}:[-1/e,0) \; \mapsto [-1,-\infty) \,$ indicates the lower real branch of the Lambert $W$ function.

The evaluation of the normalized amplitude of the noise that seeds the plasmoid instability has to be consistent with the energy spectrum. This requires to take into account the energy content at a given scale ($\delta b^2_\lambda$), as well as the probability of occurrence of a certain fluctuation amplitude on the current sheet ($f$), and finally also the projection of the fluctuation onto the unstable modes (${\delta_{\rm{in}}}/{\lambda}$). Considering these factors, the noise amplitude $\epsilon$ can be evaluated as 
\begin{equation} \label{}
\epsilon  \sim f \delta {b_\lambda} \xi \left( {\frac{\delta_{\rm{in}}}{\lambda}} \right) \, .
\end{equation}
Here, $\delta {b_\lambda} \xi$ is the magnetic flux associated with the energy content at scale $\lambda$, and its projection is obtained by multiplying it with $\delta_{\rm{in}}/{\lambda}$. The factor $f$ defines the filling fraction, and can be estimated from geometrical considerations. If we envision current sheets that form between alternately twisted flux bundles (magnetic islands), a close packed configuration yields hexagonal arrays with current sheets that develop on two of the six edges \citep[see, e.g.,][]{Zhou2014}. 
Therefore, from the area of a regular hexagon we can estimate $f \sim c_s {\lambda}/\xi$, with $c_s = 2/(3\sqrt{3})$. Finally, evaluating the inner resistive layer width  $\delta_{\rm{in}} \sim S_\xi ^{-1/4} \xi {(\lambda /\xi)^{3/4}} \sim S_\xi ^{-1/2}\xi $ as the zeroth order approximation, we obtain
\begin{equation} \label{norm_noise}
{\hat \epsilon} \sim c_s \frac{\delta_{\rm{in}}}{\xi} \sim c_s R_m^{-3/14} {\left( {\frac{\lambda_*}{\lambda}} \right)}^{\! 6/5} \, .
\end{equation}

\begin{figure}
\begin{center}
\includegraphics[width=8.6cm]{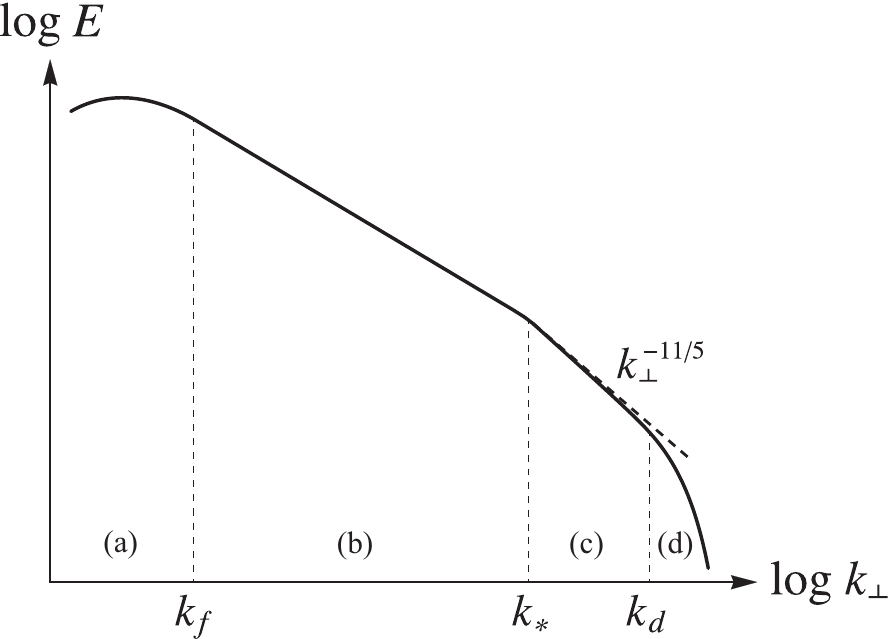}
\end{center}
\caption{Schematic diagram of the energy cascade in MHD turbulence at very large $R_m$. Labels are used to indicate the (a) energy-containing range, (b) inertial range, (c) plasmoid-mediated range, and (d) dissipation range. In the plasmoid-mediated range, the slope of the energy spectrum $E({k_\bot})$ follows Eq. (\ref{ene_spectrum}).}
\label{figSpectrum}
\end{figure}

We are now in a position to completely determine the energy spectrum in the plasmoid-mediated range. Indeed, substituting Eq. (\ref{norm_noise}) into Eq. (\ref{pert_b_2}), we have  
\begin{equation}\label{pert_b_3} 
\delta {b_\lambda} \simeq \frac{{\varepsilon ^{2/5}}{\lambda^{3/5}}}{\eta^{1/5}}{\left[ { - \frac{9}{5} W_{-1} \left( { - {{\left( \frac{\lambda_*^2}{\lambda^2 R_m^{5/14}}  \right)}^{\! 1/9}}} \right)} \right]^{2/5}} \, ,
\end{equation}
where the factor $(5/9) (2^2 c_s)^{5/3} \approx 1$ has been neglected. Then, the energy spectrum $E(k_\bot)$ can be duly calculated from the relation
\begin{equation}\label{} 
\int_{{k_ \bot }}^\infty  {E({k'_\bot})d{k'_\bot}}  \sim {(\delta {b_\lambda })}^2 \, .
\end{equation} 
Therefore, taking the derivative with respect to $k_\bot$ of this expression, we arrive at
\begin{eqnarray}\label{ene_spectrum} 
E({k_\bot}) &=& C\frac{\varepsilon^{4/5} |W_{-1}({\vartheta_k})|^{4/5}}{{\eta^{2/5} k_\bot^{11/5}}} \left[ {1 + \frac{4{k_\bot}}{5}\frac{d}{{d{k_\bot}}}\ln |W_{-1}({\vartheta_k})| } \right]  \nonumber\\
&\simeq& C\frac{{{\varepsilon ^{4/5}}}}{{{\eta ^{2/5}}}}k_ \bot ^{-11/5} |W_{-1}({\vartheta_k})|^{4/5}  \, .
\end{eqnarray}
where $C$ is a constant and 
\begin{equation}\label{} 
\vartheta_k = - {{\left( \frac{k_\bot^2}{k_*^2 R_m^{5/14}}  \right)}^{\!1/9}} \, .
\end{equation} 
Eq. (\ref{ene_spectrum}) indicates that the energy spectrum in the plasmoid-mediated range is steeper than the $-3/2$ \citep{Iroshnikov63,Kraichnan65,Boldyrev2006,MalletScheko17} and $-5/3$ \citep{Higdon84,GS95,Howes2008} slopes that are typically discussed for the standard inertial range. This is because the disruption of the current sheet structures facilitates the energy cascade towards small scales. Furthermore, differently from what was previously assumed \citep{Mallet2017,LouBold2017,BoldLou2017}, the energy spectrum is not a pure power law, as it includes also the contribution of the Lambert $W$ function. For $\vartheta_k$ approaching zero, we can consider the asymptotic expansion $W_{-1}(\vartheta_k) = \ln (-\vartheta_k) - \ln \big(-\ln(-\vartheta_k) \big) + o(1)$. In this case, keeping only the first term of this expansion, we find 
\begin{equation}\label{} 
W_{-1}(\vartheta_k) \simeq \frac{2}{9} \ln { \left( R_m^{-5/28} \frac{k_\bot}{k_*}  \right)}\, .
\end{equation} 
Therefore, from this expression we can see that the energy spectrum will be steeper than the power law component with slope $-11/5$. This situation is depicted in Fig. (\ref{figSpectrum}).

The plasmoid-mediated cascade terminates at the dissipation scale ${\lambda_d} = 2\pi /{k_d}$, which can be determined from the requirement that in steady-state the rate of energy dissipation has to be equal to the rate of energy cascade, 
\begin{align}
\frac{dE}{dt} = - \eta  \int^{k_{d}}k_{\perp}^2 E(k_{\perp}) dk_{\perp} = - \varepsilon \, ,
\end{align}
and should not depend on the magnetic Reynolds number. In practice, we have to solve the equation
\begin{align}
R_m^{-1}  \int_0^{{k_d}} {k_\bot^2} E({k_\bot}) d{k_\bot} = \frac{\varepsilon}{{V_A}L} \, ,
\end{align}
which can be expressed as
\begin{equation}\label{} 
\int_0^{{k_d}} {k_\bot^{-1/5}} |W_{-1}({\vartheta _k})|^{4/5} d{k_\bot} = \frac{{R_m^{3/5}}}{{C{L^{4/5}}}} \, .
\end{equation} 
An asymptotic approximation of this integral leads us to 
\begin{equation}\label{lambda_d} 
\frac{\lambda_d}{L} \sim C^{5/4} R_m^{-3/4} {\left|  W_{-1} {\left( {\Xi} \right)} \right| } \, ,
\end{equation}
with 
\begin{equation}\label{} 
\Xi = - \frac{9}{11} \bigg[ {\left( {\frac{4}{5C}} \right)^{\! 25/8}} \frac{R_m^{8/7}}{(k_* L)^2}\bigg]^{\!1/11} \, .
\end{equation}
Since ${k_*}L \propto R_m^{4/7}{\left| {{W_{-1}}f({R_m})} \right|^{8/7}}$, at the first order we have the approximate power-law ${\lambda_d}/L \propto R_m^{-3/4}$.

\section{Anisotropy in the plasmoid-mediated range}

As we have previously emphasized, the plasmoid instability has the effect of constraining the anisotropy of the turbulent structures by limiting their transverse aspect ratio $\xi/\lambda$. Here, we provide explicitly the scalings for the anisotropy of the fluctuating fields in the plasmoid-mediated range. They can be readily derived from the disruption condition obtained in Sec. \ref{SecCSdisr}, which can be rewritten as 
\begin{equation} \label{}
\left( {{c_\gamma }\frac{{\delta {b_\lambda }}}{\lambda }S_\lambda ^{ - 1/2}} \right)\left( {\frac{\lambda }{{\delta {b_\lambda }\sin {\theta _\lambda }}}} \right) \simeq {c_\gamma }\frac{{W(\zeta )}}{\underline \alpha} \, .
\end{equation}
Indeed, this condition yields
\begin{equation} \label{sintheta}
\sin {\theta _\lambda } \simeq  \frac{\eta^{3/5}}{\varepsilon^{1/5} \lambda^{4/5}} {\left[ {\frac{\underline \alpha}{{W(\zeta )}}} \right]^{6/5}} \propto {\left( {\frac{\lambda }{{{\lambda_*}}}} \right)^{\! - 4/5}} {\left[ {\frac{\underline \alpha}{{W(\zeta )}}} \right]^{6/5}} \, 
\end{equation}
after using Eq. (\ref{pert_b_1}) and the constant energy flux requirement at the outer-scale. The auxiliary variable $\zeta$ contains the quantities $\xi$ and $\delta {b_\lambda}$, for which we can adopt their expression at the zeroth order, since they occur in the argument of the Lambert $W$ function. In this way we have
\begin{equation}\label{} 
\zeta  =  - \frac{3}{7}{\left( {{2^6}{\hat \epsilon ^3}} \right)^{3/7}} {\left( {\frac{\lambda_*}{L}} \right)^{\! -3/7}}{\left( {\frac{\lambda}{\lambda_*}} \right)^{48/35}} \, .
\end{equation}
As can be seen from Eq. (\ref{sintheta}), the aspect ratio of the field structures in the perpendicular 2D plane, $\xi / \lambda \sim 1/{\sin {\theta_\lambda}}$, decreases towards smaller scales in the plasmoid-mediated range.  

From Eq. (\ref{sintheta}), the coherence length of the turbulent structures in the direction of the fluctuating magnetic field can be easily determined as
\begin{equation} \label{}
\frac{\xi}{L} \sim \frac{{{\varepsilon^{1/5}}{\lambda ^{9/5}}}}{{\eta ^{3/5}} L}{\left[ {\frac{{W(\zeta )}}{\underline \alpha}} \right]^{6/5}} \propto {\left( {\frac{\lambda }{{{\lambda _*}}}} \right)^{\! 9/5}}{\left[ {\frac{{W(\zeta )}}{\underline \alpha}} \right]^{6/5}} \, .
\end{equation}
Lastly, the length of the current structures, $l_\parallel$, can be calculated from the critical balance condition \citep{GS95}, which gives ${l_\parallel } \sim {V_{A}}{\tau _{{\rm{nl}}}}$. Substituting the espression for ${\tau _{{\rm{nl}}}}$ obtained in Sec. \ref{SecSpectrum} it is straightforward to obtain
\begin{equation} \label{}
\frac{l_\parallel}{L_\parallel} \sim \frac{V_A}{L_\parallel} \frac{\lambda^{6/5}}{\varepsilon^{1/5} \eta^{2/5}} {\left[ {\frac{\underline \alpha}{W(\zeta)}} \right]^{1/10}} \propto {\left( {\frac{\lambda }{{{\lambda _*}}}} \right)^{\! 6/5}}  {\left[ {\frac{\underline \alpha }{{W(\zeta )}}}  \right]^{1/10}} \, ,
\end{equation}
where we have used the constant energy flux requirement at the outer-scale.

\section{Transitional magnetic Reynolds number}\label{SecReyn}

We conclude the developed theory by determining the magnetic Reynolds number above which the plasmoid instability becomes statistically significant enough to affect the turbulent cascade before it can reach the dissipation scale. In fact, the possibility to reach the Sweet-Parker width ${\lambda} \simeq {\xi} S_\xi^{-1/2}$, which corresponds to the dissipation scale ${\lambda _\eta}$ in the absence of plasmoids, depends on the value of the magnetic Reynolds number of the system under consideration \citep{Comisso2016,Comisso2017,Huang2017}.

For Alfv{\'e}nic current sheet formation, it was shown that a current sheet disrupts when its aspect ratio is smaller than the Sweet-Parker one if its Lundquist number is greater than the ``transitional'' Lundquist number \citep{Comisso2017} 
\begin{equation}\label{S_transit} 
S_{\xi_T} = {\left[ {{\tilde \alpha} \,W\left( {\frac{1}{{\tilde \alpha }}{{\left( {\frac{1}{{{2^6}{\hat \epsilon ^3}}}} \right)}^{\! 1/\tilde \alpha}}} \right)} \right]^4} \, ,
\end{equation}
where $\tilde \alpha = 2(4-\alpha)$. From this expression, using $\alpha=4/3$ and recalling that $R_{m_T} \sim (L/{\lambda_*}) S_{\xi_T}$, we can obtain
\begin{equation}\label{Rm_transit} 
R_{m_T} \sim S_{{\xi _T}}^{7/3}{\left[ { - \frac{7}{3}\,{W_{-1}}\left( { - \frac{3}{7}{{\left( {{2^6}{\hat \epsilon ^3}} \right)}^{3/7}}S_{\xi_T}^{4/7}} \right)} \right]^{8/3}} \, .
\end{equation}
Therefore, using Eq. (\ref{S_transit}) and $\hat \epsilon \sim c_s R_{m_T}^{-3/14}$ at $\lambda = \lambda_*$, we find that plasmoid formation becomes sufficiently important to modify the dissipation scale and the near-dissipation part of the inertial range when
\begin{equation}\label{} 
R_m > R_{m_T} \sim 5 \times 10^6 \, .
\end{equation}
This is a very modest value of the magnetic Reynolds number for many of the astrophysical systems where MHD turbulence is thought to play an essential role, such as the solar corona, black hole accretion disks, the interstellar medium, galaxies and galaxy clusters. Therefore, plasmoids are expected to modify the turbulent path to dissipation in these systems.

\section{Conclusions} \label{SecConc}

In this paper, we have formulated a self-consistent theory of MHD turbulence in a regime where the turbulent structures are unstable to the formation of plasmoids via magnetic reconnection. A distinctive feature of this theory is that it accounts for the mutual interplay between turbulence and plasmoid formation, which is found to be important when $R_m > R_{m_T}$.  
Following the theory of the plasmoid instability developed in \citet{Comisso2016,Comisso2017}, we have shown how the fluctuations arising from turbulence provide the noise that seeds the plasmoid growth, and we have determined when the plasmoids break up the turbulent structures in which they grow. The disruption of these sheet-like structures leads to a modification of the turbulent energy cascade, which, in turn, exerts a feedback effect on the plasmoid instability via the noise generated by the turbulence itself. 

We find that the standard inertial range of the turbulent cascade terminates at the length scale $\lambda_*$ given by Eq. (\ref{lambda_*}), with $\alpha=4/3$ and ${\hat \epsilon} \sim c_s R_m^{-3/14}$. Below this scale the current sheet structures that form in the turbulent environment are disrupted by the plasmoid instability, which has the effect of steepening the energy spectrum. In this plasmoid-mediated range, we find that the energy spectrum follows Eq. (\ref{ene_spectrum}), namely $E({k_\bot}) \propto k_ \bot ^{-11/5} |W_{-1}({\vartheta_k})|^{4/5}$, which turns out to be steeper than the power law factor with slope $-11/5$. The aspect ratio $\xi/\lambda \sim 1/{\sin {\theta_\lambda}}$ of the sheet-like structures in the plasmoid-mediated range decreases  towards smaller scales as indicated in Eq. (\ref{sintheta}). Finally, the dissipation scale is reached at the length scale $\lambda_d$ given by Eq. (\ref{lambda_d}). 

As a consequence of the complex interaction between the turbulence dynamics and the magnetic reconnection process that occurs in the turbulent environment, we find that the scaling relations of the turbulent cascade are not true power laws, which is a result that has never been derived before. At the zeroth order we reproduce the power laws obtained by \citet{Mallet2017} and \citet{BoldLou2017}, but more accurate relations are required to make quantitative predictions. For example, the neglect of the factor $\left( {{\bar \alpha}/{W(\chi)}} \right)^{8/7}$ in Eq. (\ref{lambda_*}) would overestimate the length scale $\lambda_*$ by one order of magnitude or more, depending on the astrophysical system under consideration. Therefore, future studies can gainfully employ the obtained scaling relations to evaluate the effects of MHD turbulence in astrophysical systems where it plays a fundamental role, such as stellar coronae and accretion disks.

\acknowledgments
It is a pleasure to acknowledge fruitful discussions with Yuri Cavecchi, Fatima Ebrahimi, Russell Kulsrud, and Takuya Shibayama. This research was supported by the NSF Grant Nos. AGS-1338944 and AGS-1460169 and by DOE Grant No. DE-AC02-09CH-11466.

\end{document}